\documentclass[twocolumn]{jjap2}
\usepackage{txfonts}

\title{Sensitive label-free IgG detection using MEMS QCM biosensor with 125-MHz wireless quartz resonator}

\author{Lianjie Zhou$^{1}$, Fumihito Kato$^{2}$, and Hirotsugu Ogi$^{1}$\thanks{E-mail: ogi@prec.eng.osaka-u.ac.jp}}

\inst{
$^{1}$Graduate School of Engineering, Osaka University, Yamadaoka 2-1, Suita, Osaka 565-0871, Japan\\
$^{2}$Department of Mechanical Engineering, Nippon Institute of Technology, Saitama 345-8501, Japan\\
}

\abst{We present a wireless QCM biosensor fabricated using MEMS technology. The MEMS QCM biosensor contains a 125-MHz AT-cut quartz resonator embedded in the microchannel. Because of the compact design, the MEMS QCM biosensor is suitable for mass production and device miniaturization. We performed the IgG detection measurement with different concentrations. The detection limit was l ng/mL or less, which is superior to that of the gold-standard SPR method. Furthermore, we studied the binding affinity between protein A and IgG by studying the frequency response of the QCM biosensor.  It shows good agreement with reported values. Therefore, the presented MEMS QCM biosensor has advantages of compact, low cost, low power consumption, high sensitivity, and reliability.   
\\
\\
Japanese Journal of Applied Physics 60, SDDB03 (2021)\\
https://doi.org/10.35848/1347-4065/abea50
}

\begin{document}
\maketitle

\section{Introduction}
Owing to advantages of high stability, simple instrumentation, label free, and capability of real-time monitoring, the quartz crystal resonator has been applied in a variety of areas, such as, frequency source of communication systems\cite{Y.Watanabe}, gas sensors\cite{P.Si, M.Vagra, L.Zhou, K.W.Liu}, biosensors\cite{H.J.Lim, D.Milioni, D.Chen, K.Noi2, N.Asai}, and so on. Based on the mass loading effect, quartz crystal microbalance (QCM) coated with water-absorbing materials, including graphene oxides\cite{S.Lee}, nanodiamond film \cite{Y.Yao}, TiO$_{2}$ \cite{D.Zhang-1}, SnO$_{2}$ nanowires \cite{N.Gao}, and polydopamine/graphene oxide \cite{D.Zhang-2} have been developed as sensitive humidity sensors. In addition, metal oxide nanofibrous membrane \cite{N.Horzum}, monolayer graphene film \cite{V.V.Quang}, manganese oxide nanosheets \cite{Y.Tokura}, and carbon nanotubes \cite{E.S.Manoso} have also been deposited on QCM to fabricate sensors for detecting volatile organic compounds. Among various biosensors, the electrochemical and surface plasma resonance (SPR) biosensors enable real-time and label-free detection with high sensitivity \cite{K.Tsugimura, S.Kasturi, T.Springer, M.Piliarik, J.Homola}. The QCM biosensor can also monitor binding reaction between ligand and target molecules, providing useful information, such as the concentration of the target and the binding affinity between two molecules. Suthar et al. described a sensitive and selective detection of specific exosomes in human serum using QCM without the need of a secondary label \cite{J.Suthar}. Combining QCM with aptamer-functionalized gold nanoparticles, Chen et al. achieved a detection limit of 0.1 nM for human $\alpha$-thrombin \cite{Q.Chen}. Guha et al. succeed in detecting specific biomarker of bacterial infection with concentration of 1 $\mu$M using a QCM coated with nanomolecular imprinted polymer \cite{A.Guha}. Furthermore, the viscoelastic property or structural change of the attached protein can be evaluated by QCM \cite{T.Shagawa}. Ogi et al. and Hamada et al. monitored the deposition reaction and fibrillation procedure of amyloid $\beta$ peptides using the QCM and internal-reflection-fluorescence microscopy (QCM-TIRFM) system \cite{H.Ogi-SR, H.Hamada}. Shoaib et al. introduced a quartz crystal microbalance with dissipation (QCM-D) sensing strategy to assess the changes in the viscoelastic properties of a cell monolayer in response to oxidative stress \cite{S.Shoaib}. Muramatsu et al. investigated the attachment process and morphological changes of cells through the frequency change of QCM \cite{H.Muramatsu}. However, in a conventional QCM, metal thin films are deposited on both sides of the quartz resonator as electrodes. In addition, mechanical contacts are necessary between the electrodes and the electric connectors. This configuration imposes restriction on the available frequency and deteriorates the quality factor of QCM. 

To improve the performance of the QCM biosensor, the wireless and electrodeless QCM has been developed\cite{H.Ogi-AnalChem1, H.Ogi-Proc.Jpn.Acad}. It operates in a wireless manner, allowing the use of extremely thin and high frequency resonator. Because of the higher mass sensitivity of thinner quartz resonator, the  QCM sensitivity can be greatly improved using high frequency and wireless quartz resonator. Noi et al. performed the mass-amplified sandwich assay with bio-nanocapsules (BNCs) using the wireless multichannel QCM biosensor and achieved a detection limit of 10 pg/mL for the detection of C-reactive protein \cite{K.Noi}. However, the thin quartz resonator need to be manually set in the sensor cell, and such subtle and time-consuming procedure deteriorates the practicability. 

In our previous study, we demonstrated the QCM biosensor with a naked embedded wireless quartz resonator fabricated using Micro-Electro-Mechanical Systems (MEMS) technology\cite{F.Kato-BB}.  A long fluid microchannel was adopted there in order to obtain steady solution flow and stable baseline, resulting in a large sensor size (20 mm in length). In this study, we present a new MEMS QCM biosensor packaged with a wireless quartz resonator. The length of fluid microchannel has been reduced by half, accordingly, the fabricated MEMS QCM biosensor becomes more compact. Importantly, the compact QCM sensor chip is suitable for mass production and beneficial to device miniaturization. Moreover, the shorter microfluidic channel makes it possible to perform the bioassay with minimal consumption of analyte, which is important especially in diagnostic application. For demonstrating the high sensitivity and usefulness of the MEMS QCM biosensor, we performed the immunoglobulin G (IgG) detection measurement with different concentrations. We then evaluated the affinity between protein A and IgG by studying the frequency response during the binding reaction.

\section{Experiments}
\subsection{Fabrication of MEMS QCM biosensor}

Figure 1(a) illustrates the structure of MEMS QCM biosensor. It consists of glass substrates, the AT-cut quartz resonator, and silicon substrate. Microchannels were constructed on the upper glass substrate and the silicon layer using the MEMS technology similar to the process we previously developed\cite{F. Kato-BB, F.Kato-JJAP}. Here, we briefly explain the procedure: For fabricating the upper glass substrate, the resist mask was patterned on one side of the glass wafer. The isotropic etching was then performed to construct the microchannel and micropillars in the glass wafer. Subsequently, the mask for the inlet and outlet ports were patterned by photolithography on the dry film photoresist mask followed by the sandblast process. Similarly, the microchannel and micropillars were fabricated on the silicon layer using the inductively coupled plasma-reactive ion etching (ICP-RIE) method. The microchannel is 30 $\mu$m in depth, containing a 1.85 $\times$ 1.65 mm$^{2}$ area for embedding the resonator as shown in Fig. 1(b). Micropillars and semi-cylindrical sidewall were built to support and confine the resonator inside the microchannel. The  resonator has an in-plane area of 1.8$\times$1.6 mm$^{2}$. The thickness is 13.5 $\mu$m, corresponding to the fundamental thickness-shear resonant frequency of 125 MHz. To fabricate the quartz resonator, an AT-cut quartz wafer which was bonded to a silicon wafer was first polished to certain thickness, and it was cut into chips with smaller dimensions. The silicon substrate was finally removed by a wet process. Thin films of 2 nm chromium and 8 nm gold were deposited on both sides of the resonator using the radio frequency (RF) magnetron-sputtering method. After inserting the QCM, we packaged the sensor chip using the anodic bonding method. Figure 1(c) illustrates the fabricated MEMS QCM biosensor. The dimension of the sensor chip is 10$\times$5$\times$0.6 mm$^3$. The Q value of the QCM exceeds 80000 in air, but it is about 300 in solution, which is similar with our previously reported value\cite{H.Ogi-AnalChem2}. Figure 2 demonstrates the mass production procedure of many identical MEMS QCM biosensors. Because of the miniaturization we are able to fabricate about 200 identical sensor chips simultaneously using the three pieces of 4 inch wafers.
\begin{figure}
\begin{center}
\includegraphics[width=85mm]{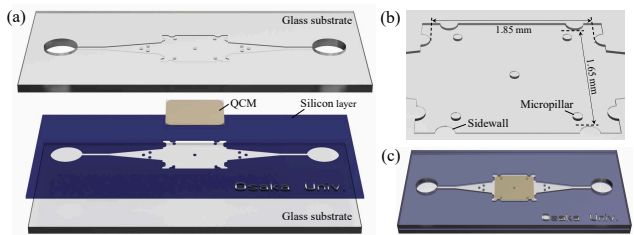}
\end{center}
\caption{(a) Components of the MEMS QCM biosensor. (b) Structure of the microchannel. (c) Fabricated MEMS QCM biosensor.}
\label{Fig1Structure}
\end{figure}

\begin{figure*}[t]
\begin{center}
\includegraphics[width=110mm]{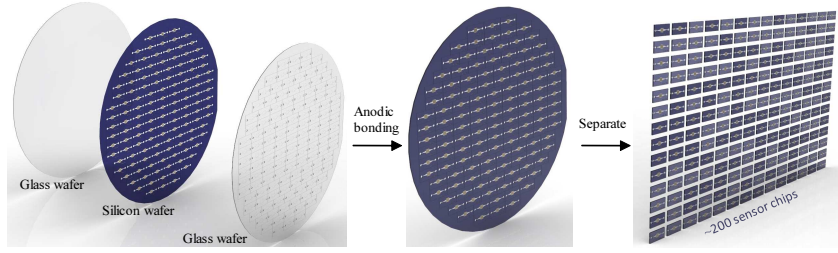}
\end{center}
\caption{Mass production process of the MEMS QCM biosensor. }
\label{Fig2MassProduction}
\end{figure*}

\subsection{Functionalization of QCM surface for IgG detection}
Figure 3 describes the process of QCM surface functionalization and the IgG detection\cite{USE-Zhou}. The QCM was first cleaned with a piranha solution (98\% H$_{2}$SO$_{4}$: 30\% H$_{2}$O$_{2}$ = 7:3) and then rinsed with ultrapure water. To make the self-assembled monolayer (SAM), we injected 10 mM SAM molecules (10-carboxy-decanthiol) in absolute ethanol and incubated overnight at 4 $^{\circ}$C. After rinsing with absolute ethanol and ultrapure water,  a mixture of 100 mM N-hydroxysulfosuccinimide sodium salt (NHS) and 100 mM 1-Ethyl-3-(3-dimethylaminopropyl) carbodiimide hydrochloride (EDC) in ultrapure water was injected, followed by incubation at room temperature for 1 h to activate the SAM terminals. After rinsing with ultrapure water, a 200 $\mu$g/mL protein A in phosphate buffer solution (PBS) was injected. The sensor chip was then incubated for 2 h at room temperature and rinsed by the PBS buffer. At last, a 5 mg/mL bovine serum albumin (BSA) in PBS was injected to block the remaining activated terminals. In the IgG detection measurement, we prepared rabbit IgG (rIgG) in ultrapure water with various concentrations (1--1000 ng/mL). The detection was repeated after dissociating the IgG molecules from QCM surface by flowing glycine-HCl buffer (GHB) solution.

\begin{figure*}[t]
\begin{center}
\includegraphics[width=120mm]{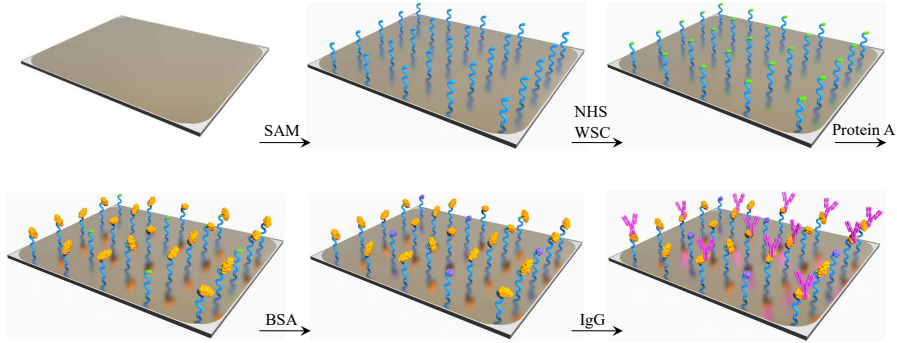}
\end{center}
\caption{Schematic of the QCM surface functionalization and IgG detection.}
\label{Fig3Functionalization}
\end{figure*}

\begin{figure}[b]
\begin{center}
\includegraphics[width=85mm]{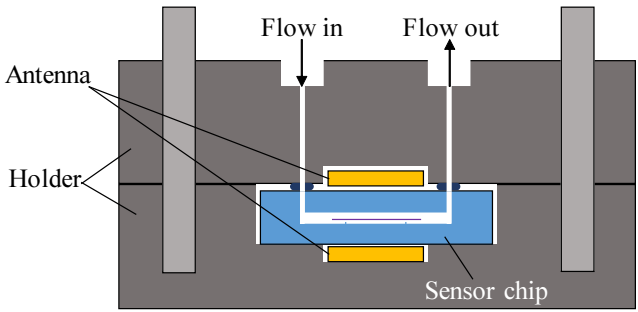}
\end{center}
\caption{Schematic of the sensor cell.}
\label{Fig4SensorCell}
\end{figure}

\subsection{Materials and reagents}
10-carboxy-decanthiol (\#C385) and EDC (\#W001) were purchased from DOJINDO. NHS (\#087-09371), rIgG (\#148-09551), GHB (0.1M, pH2.2) and PBS (0.1M, pH6.8) were purchased from Wako. Protein A (\#101100) was purchased from Invitrogen. BSA (\#A3059) was purchased from Sigma Aldrich. 
\subsection{Experiment system}
In the IgG detection measurement, the sensor chip was set in the homebuilt sensor cell as shown in Fig. 4. The excitation of QCM and signal readout were performed by the line antennas located outside the sensor chip. The line antennas were made of copper on the printed circuit board (PCB); they are 2 mm in width and 35 mm in length. The resonant frequency was monitored by a network analyzer (ZNLE3, Rohde \& Schwarz). A micro-pump was used to flow and circulate the buffer and analyte solutions with a flow rate of 400 $\mu$L/min.

\begin{figure}[b]
\begin{center}
\includegraphics[width=80mm]{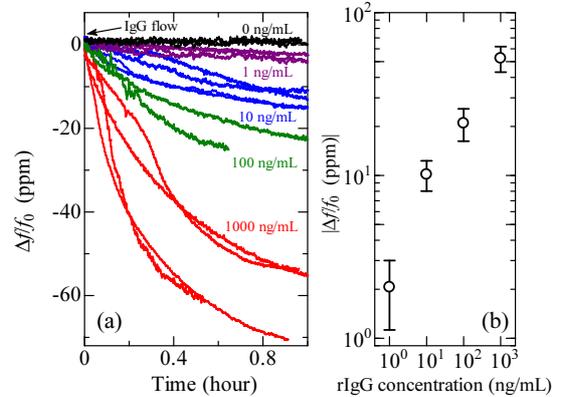}
\end{center}
\caption{(a) Frequency responses measured by the MEMS QCM biosensor during the IgG-solution flow. (b) Relationship between the amount of the frequency decrease at 35 min from the injection and the IgG concentration.}
\label{Fig5IgGdetection}
\end{figure}

\section{Results and Discussion} 
\subsection{IgG detection} 
Using the MEMS QCM biosensor, we performed IgG detection with concentration from 1 to 1000 ng/mL. Figure 5(a) shows the frequency responses of the QCM biosensor during the binding reaction between IgG and protein A with different IgG concentrations; after a binding reaction, the IgG molecules were removed by injecting the GHB solution.  After flowing an IgG solution, the resonance frequency starts to decrease as the binding reaction proceeds. Even for the injection of 1 ng/mL IgG, obvious frequency decrease can be observed. The QCM biosensor exhibited lager decreases in frequency  for IgG solutions with higher concentrations, indicating the good reusability of the QCM biosensor. The relationship between the frequency decrease after 35 min from the injection and the IgG concentration is shown in Fig. 5(b), showing a good correlation. Furthermore, we performed the equivalent IgG detection measurement using the surface plasma resonance (SPR) system (Biacore 3000).  The SPR biosensor showed no response to the 1 ng/mL IgG solution and displayed a detection limit of 5 ng/mL, confirming sufficiently higher sensitivity of our newly developed MEMS biosensor.

 In the manual functionalization of QCM surface, the areal density of immobilized SAM molecules and protein A vary with the amount of injected solution and incubation time, resulting in the dispersion of frequency response for higher concentration IgG (1000 ng/mL). Therefore, although without optimization of immobilization condition, the MEMS QCM biosensor shows high sensitivity in IgG detection with a detection limit of 1 ng/mL or lower. 

\subsection{Binding affinity evaluation} 
We also investigated the binding affinity between IgG and protein A by studying the frequency response of the MEMS QCM biosensor. Assuming the pseudo-first-order reaction between two biomolecules, the frequency of QCM decreases exponentially when the binding reaction proceeds\cite{Y.Liu, H.Ogi-BB}, 
\begin{equation}
\label{freqLinear}
\frac{\Delta f(t)}{f_{0}}=A\left \{ e^{-(k_{a}C_{IgG}+k_{d})t}-1 \right \}
\end{equation} 
with the IgG concentration $C_{IgG}$, the reaction velocity constants for association $k_{a}$ and dissociation $k_{d}$. The equilibrium constant which represents the binding affinity is defined as $K_{A}=k_{a}/k_{d}$. We determined the exponential coefficient $\alpha$ during binding reaction in Fig. 5(a). Fig. 6 shows the relationship between the exponential coefficient and the IgG concentration. The determined values of the kinetics constants are $k_{a}=1.71\times 10^{5}(M^{-1}s^{-1})$, $k_{d}=5.35\times 10^{-4}(s^{-1})$ and $K_{A}=3.20\times 10^{8}(M^{-1})$. The binding affinity between protein A and IgG obtained in this study is in good agreement with reported values in literature (1.59--3.45$\times 10^{8}M^{-1}$(rIgG)\cite{S.C.Kuo}, 1.4$\times 10^{8}M^{-1}$(hIgG)\cite{H.G.Svensson} and 5.54$\times 10^{8}M^{-1}$(hIgG)\cite{K.Noi-JJAP}).

\begin{figure}[t]
\begin{center}
\includegraphics[width=80mm]{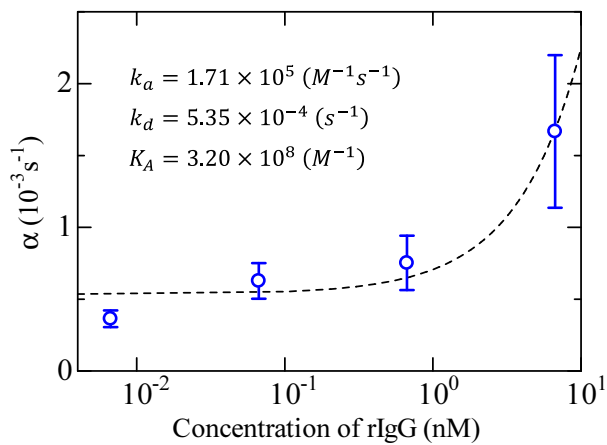}
\end{center}
\caption{Relationship between the exponential coefficient of frequency decrease and the IgG concentration.  The dashed line indicates the fitted linear function.}
\label{Fig6Affinity}
\end{figure}

\section{Conclusion} 
We presented a MEMS biosensor with a 125-MHz wireless QCM, where a thin quartz resonator was embedded in the microchannel and supported by micropillars. The MEMS QCM biosensor has advantages of compact, low cost, reusable and high sensitivity. Furthermore, a multitude of sensor chips were fabricated simultaneously using the MEMS technology. We functionalized the MEMS QCM biosensor by immobilizing the SAM molecules and protein A on the QCM surface and performed the IgG detection measurements with various concentrations. The MEMS QCM biosensor shows high sensitivity, which succeeded in detecting IgG with concentration as low as 1 ng/mL. We studied the binding affinity between the rIgG and protein A. An affinity of $K_{A}=3.20\times 10^{8}(M^{-1})$ was obtained, which is in good agreement with reported values, indicating the good reliability of the MEMS QCM biosensor. However, because the QCM operates wirelessly by antenna, the signal deteriorates when the QCM is surrounded by solution which contains a large amount of charge carriers. Improvement of signal strength will be our further study.

\section*{Acknowledgement}
This research was supported by Development of Advanced Measurement and Analysis Systems from Japan Science and Technology Agency (Project No. JP-MJSN16B5).

\end{document}